\documentclass[letterpaper,12pt,nofootinbib,aps,tightenlines,superscriptaddress]{revtex4}
\usepackage{fancyhea}
\usepackage{graphicx}
\usepackage{hyperref}      
\usepackage[usenames,dvipsnames]{color} 
\usepackage{amssymb}
\usepackage{amsmath}
\usepackage{titlesec}

\usepackage{natbib,hyperref,ifthen}

\newcommand{\ANLHEP}{HEP Division, Argonne National Laboratory, Lemont, IL 60439, USA}

\newcommand{\ASU}{Arizona State University, Tempe, AZ  85287}

\newcommand{\BenGurion}{Department of Physics, Ben-Gurion University, Be'er Sheva 84105, Israel}

\newcommand{\BNL}{Brookhaven National Laboratory, Upton, NY 11973}

\newcommand{\Caltech}{California Institute of Technology, Pasadena, CA 91125}

\newcommand{\CCA}{Center for Computational Astrophysics, 162 5th Ave, 10010, New York, NY, USA}

\newcommand{\CPPM}{Aix Marseille Univ, CNRS/IN2P3, CPPM, Marseille, France}

\newcommand{\CITA}{Canadian Institute for Theoretical Astrophysics, University of Toronto, Toronto, ON M5S 3H8, Canada}

\newcommand{\CMUCosmo}{Department of Physics, McWilliams Center for Cosmology, Carnegie Mellon University}

\newcommand{\Cornell}{Cornell University, Ithaca, NY 14853}

\newcommand{\damtp}{DAMTP, Centre for Mathematical Sciences, Wilberforce Road, Cambridge, UK, CB3 0WA}

\newcommand{\dunlap}{Dunlap Institute for Astronomy and Astrophysics, University of Toronto, ON, M5S3H4}

\newcommand{\EPFL}{Institute of Physics, Laboratory of Astrophysics, Ecole Polytechnique Fédérale de Lausanne (EPFL), Observatoire de Sauverny, 1290 Versoix, Switzerland}

\newcommand{\FNAL}{Fermi National Accelerator Laboratory, Batavia, IL 60510}

\newcommand{\FQAUB}{Dept. de F\' isica Qu\` antica i Astrof\' isica, Universitat de Barcelona, Mart\' i i Franqu\` es 1, E08028 Barcelona, Spain}

\newcommand{\GRAPPA}{GRAPPA Institute, University of Amsterdam, Science Park 904, 1098 XH Amsterdam, The Netherlands}

\newcommand{\GSFC}{Goddard Space Flight Center, Greenbelt, MD 20771 USA}

\newcommand{\HarvardPhys}{Department of Physics, Harvard University, Cambridge, MA 02138, USA}

\newcommand{\Heidelberg}{Max Planck Institute for Astronomy, Heidelberg, Germany}

\newcommand{\ICC}{ICC, University of Barcelona, IEEC-UB, Mart\' i i Franqu\` es, 1, E08028 Barcelona, Spain}

\newcommand{\ICE}{Institute of Space Sciences (ICE, CSIC), Campus UAB, Carrer de Can Magrans, s/n, 08193 Barcelona, Spain}

\newcommand{\ITFA}{Institute for Theoretical Physics, University of Amsterdam, Science Park 904, 1098 XH Amsterdam, The Netherlands}

\newcommand{\JHU}{Johns Hopkins University, Baltimore, MD 21218}

\newcommand{\JPL}{Jet Propulsion Laboratory, California Institute of Technology, Pasadena, CA, USA}

\newcommand{\KASSI}{Korea Astronomy and Space Science Institute, Daejeon 34055, Korea}

\newcommand{\kavli}{Kavli Institute for Cosmology, Cambridge, UK, CB3 0HA}

\newcommand{\KICP}{Kavli Institute for Cosmological Physics, Chicago, IL 60637}

\newcommand{\KIPAC}{Kavli Institute for Particle Astrophysics and Cosmology, Stanford 94305}

\newcommand{\LBL}{Lawrence Berkeley National Laboratory, Berkeley, CA 94720}

\newcommand{\McGill}{McGill University, Montreal, QC H3A 2T8, Canada}

\newcommand{\MIT}{Massachusetts Institute of Technology, Cambridge, MA 02139}

\newcommand{\Marseille}{Laboratoire d'Astrophysique de Marseille, Aix Marseille Univ, CNRS, CNES, LAM, Marseille, France}

\newcommand{\Missouri}{Department of Physics, Missouri University of Science and Technology, 1315 N Pine St., Rolla, MO 65409, USA}

\newcommand{\NAOC}{National Astronomical Observatories, Chinese Academy of Sciences, PR China}

\newcommand{\NCBJ}{National Center for Nuclear Research, Ul.Pasteura 7,Warsaw, Poland}

\newcommand{\NOAO}{National Optical Astronomy Observatory, 950 N. Cherry Ave., Tucson, AZ 85719 USA}

\newcommand{\NYU}{New York University, New York, NY 10003}

\newcommand{\OSU}{The Ohio State University, Columbus, OH 43212}

\newcommand{\OU}{Department of Physics and Astronomy, Ohio University, Clippinger Labs, Athens, OH 45701, USA}

\newcommand{\Oslo}{Institute of Theoretical Astrophysics, University of Oslo,  Svein Rosselands hus, Sem Sælands vei 130371 OSLO, Norway}

\newcommand{\Oxford}{The University of Oxford, Oxford OX1 3RH, UK}

\newcommand{\PI}{Perimeter Institute, Waterloo, Ontario N2L 2Y5, Canada}

\newcommand{\Pisa}{Scuola Normale Superiore, Piazza dei Cavalieri 7, 50126, Pisa, Italy}

\newcommand{\Port}{Institute of Cosmology \& Gravitation, University of Portsmouth, Dennis Sciama Building, Burnaby Road, Portsmouth PO1 3FX, UK}

\newcommand{\Princeton}{Princeton University, Princeton, NJ 08544}

\newcommand{\QMUL}{Queen Mary University of London, Mile End Road, London E1 4NS, United Kingdom}

\newcommand{\SCIPP}{University of California at Santa Cruz, Santa Cruz, CA 95064}

\newcommand{\Sejong}{Department of Physics and Astronomy, Sejong University, Seoul, 143-747, Korea}

\newcommand{\Sinica}{Academia Sinica Institute of Astronomy and Astrophysics, 645 N. A'ohoku Place, Hilo, HI 96720, USA}

\newcommand{\Stanford}{Stanford University, Stanford, CA 94305}

\newcommand{\StonyBrook}{Stony Brook University, Stony Brook, NY 11794}

\newcommand{\SussexAstronomy}{Astronomy Centre, School of Mathematical and Physical Sciences, University of Sussex, Brighton BN1 9QH, United Kingdom}

\newcommand{\Tsinghua}{Department of Physics and Tsinghua Center for Astrophysics, Tsinghua University, Beijing 100084, P R China}

\newcommand{\UAS}{Steward Observatory, University of Arizona, Tucson, AZ  85721}

\newcommand{\UCB}{Department of Astronomy, University of California Berkeley, Berkeley, CA 94720, USA}

\newcommand{\UCBP}{Department of Physics, University of California Berkeley, Berkeley, CA 94720, USA}

\newcommand{\UCBSSL}{Space Sciences Laboratory, University of California Berkeley, Berkeley, CA 94720, USA}

\newcommand{\UChicago}{University of Chicago, Chicago, IL 60637}

\newcommand{\UCI}{University of California, Irvine, CA 92697}

\newcommand{\UCLA}{University of California at Los Angeles, Los Angeles,  CA 90095}

\newcommand{\UCL}{University College London, WC1E 6BT London, United Kingdom}

\newcommand{\UFL}{University of Florida, Gainesville, FL 32611}

\newcommand{\UGTO}{Divisi\'on de Ciencias e Ingenier\'ias, Universidad de Guanajuato, Le\'on 37150, M\'exico}

\newcommand{\UNIPD}{Dipartimento di Fisica e Astronomia ``G. Galilei'',Universit\`a degli Studi di Padova, via Marzolo 8, I-35131, Padova, Italy}

\newcommand{\UPenn}{Department of Physics and Astronomy, University of Pennsylvania, Philadelphia, Pennsylvania 19104, USA}

\newcommand{\UR}{Department of Physics and Astronomy, University of Rochester, 500 Joseph C. Wilson Boulevard, Rochester, NY 14627, USA}

\newcommand{\UWMadison}{Department of Physics, University of Wisconsin - Madison, Madison, WI 53706}

\newcommand{\UWC}{Department of Physics \& Astronomy, University of the Western Cape, Cape Town 7535, South Africa}

\newcommand{\VSI}{Van Swinderen Institute for Particle Physics and Gravity, University of Groningen, Nijenborgh 4, 9747~AG~Groningen, The~Netherlands}

\newcommand{\WVU}{CSEE, West Virginia University, Morgantown, WV 26505, USA}

\newcommand{\WVUGWAC}{Center for Gravitational Waves and Cosmology, West Virginia University, Morgantown, WV 26505, USA}

\newcommand{\YorkU}{Department of Physics and Astronomy, York University, Toronto, Ontario M3J 1P3, Canada}

\usepackage[normalem]{ulem}
\usepackage{color}

\def\lsim{\raise-.75ex\hbox{$\buildrel<\over\sim$}}

\begin{document}

\title{ Astro2020 Science White Paper: \\  Astrophysics and Cosmology with Line-Intensity Mapping} 

\author{Ely D. Kovetz}
\email{kovetz@bgu.ac.il; Phone: +972-545953349}
\affiliation{\footnotesize \vspace{-0.05in} Department of Physics, Ben-Gurion University, Be'er Sheva 84105, Israel}
\affiliation{\footnotesize \vspace{-0.05in} Department of Physics and Astronomy, Johns Hopkins University, 3400 N.\ Charles St., Baltimore, MD 21218, USA}
\author{Patrick C. Breysse}
\email{pcbreysse@cita.utoronto.ca}
\affiliation{\footnotesize \vspace{-0.05in} Canadian Institute for Theoretical Astrophysics, University of Toronto, 60 George St., Toronto, ON, M5S 3H8, Canada}
\author{Adam Lidz}
\email{alidz@sas.upenn.edu}
\affiliation{\footnotesize \vspace{-0.05in} Department of Physics and Astronomy, University of Pennsylvania, 209 South 33rd Street, Philadelphia, PA 19104, USA}
\author{Jamie Bock}
\affiliation{\footnotesize \vspace{-0.05in} Division of Physics, Math and Astronomy, California Institute of Technology, 1200 E.\ California Blvd. Pasadena, CA 91125}
\author{Charles M.~Bradford}
\affiliation{\footnotesize \vspace{-0.05in} Division of Physics, Math and Astronomy, California Institute of Technology, 1200 E.\ California Blvd. Pasadena, CA 91125}
\author{Tzu-Ching Chang}
\affiliation{\footnotesize \vspace{-0.05in} Jet Propulsion Laboratory, California Institute of Technology, Pasadena, CA 91109}
\author{Simon Foreman}
\affiliation{\footnotesize \vspace{-0.05in} Canadian Institute for Theoretical Astrophysics, University of Toronto, 60 George St., Toronto, ON, M5S 3H8, Canada}
\author{Hamsa Padmanabhan}
\affiliation{\footnotesize \vspace{-0.05in} Canadian Institute for Theoretical Astrophysics, University of Toronto, 60 George St., Toronto, ON, M5S 3H8, Canada}
\author{Anthony Pullen}
\affiliation{\footnotesize \vspace{-0.05in} Center for Cosmology and Particle Physics, Department of Physics, New York University, New York, NY, 10003, USA}
\author{Dominik Riechers}
\affiliation{\footnotesize \vspace{-0.05in} Department of Astronomy, Cornell University, Ithaca, NY 14853, USA}
\author{Marta B.~Silva}
\affiliation{\footnotesize \vspace{-0.05in} Institute of Theoretical Astrophysics, University of Oslo, P.O. Box 1029 Blindern, N-0315 Oslo, Norway}
\author{Eric Switzer}
\affiliation{\footnotesize NASA Goddard Space Flight Center, Greenbelt, MD, USA}

\begin{abstract}


Line-Intensity Mapping is an emerging technique which promises new insights into the evolution of the Universe, from star formation at low redshifts to the epoch of reionization and cosmic dawn. It measures the integrated emission of atomic and molecular spectral lines from galaxies and the intergalactic medium over a broad range of frequencies, using instruments with aperture requirements that are greatly relaxed relative to surveys for single objects.  A coordinated, comprehensive, multi-line intensity-mapping experimental effort can efficiently probe over $80\%$ of the volume of the observable Universe - a feat beyond the reach of other methods. 

Line-intensity mapping will uniquely address a wide array of pressing mysteries in galaxy evolution, cosmology, and fundamental physics. Among them are the cosmic history of star formation and galaxy evolution, the compositions of the interstellar and intergalactic media, the physical processes that take place during the epoch of reionization, cosmological inflation, the validity of Einstein's gravity theory on the largest scales, the nature of dark energy and the origin of dark matter.

\end{abstract}


\maketitle

\vspace{-0.2in}

{\small \bf Thematic Science Areas: Galaxy Evolution, Cosmology and Fundamental Physics} 








\newpage

\eject
\pagenumbering{arabic} 
\setcounter{page}{1}

\section{Introduction}
\vspace{-0.25in}

Line-intensity mapping (LIM) \cite{Kovetz:2017agg} measures the spatial fluctuations in the integrated emission from spectral lines originating from many individually unresolved galaxies and the diffuse IGM to track the growth and evolution of cosmic structure. Line fluctuations trace the underlying large-scale structure of the Universe, while the frequency dependence can be used to measure the redshift distribution of the line emission along the line of sight. Traditional galaxy surveys probe discrete objects whose emission is bright enough to be imaged directly. {\em LIM is advantageous as it is sensitive to all
sources of emission in the line and thus enables the universal study of galaxy formation and evolution}. As high angular resolution is not required, {\em LIM can cover large sky areas in a limited observing time, allowing various tests of the standard cosmological model, and beyond it, across under-explored volumes of the observable Universe.} In addition, relaxed angular resolution requirements are an important attribute for space-borne instruments, where aperture drives cost, but low photon backgrounds yield very high surface brightness sensitivity.

To illustrate the promise of LIM, consider as a figure-of-merit the number $N_{\rm modes}$ of accessible modes. As the uncertainty on any quantity we wish to measure roughly scales as $1/\sqrt{N_{\rm modes}}$, the goal is clearly to maximize this number. The cosmic microwave background (CMB), which provides the farthest observable accessible to measurement, contains $N_{\rm modes}\sim\ell^2_{\rm max}\sim10^7$ modes. 
Intensity mapping of a chosen line at a given frequency provides maps that resemble the CMB, but with two important advantages: (i) there is no diffusion (Silk) damping, so small scale information can in principle be harvested down to the Jeans scale; (ii) huge redshift volumes can be measured in tomography through hyperspectral mapping. The total number of modes, $N_{\rm modes}^{\rm LIM}\sim\ell_{\rm max}^2\times N_z$, can potentially reach as high as $10^{16}(!)$ \cite{Loeb:2003ya}, limited in reality by partial sky coverage and both diffuse and line foreground contamination. Compared to galaxy surveys, LIM retains full spectral resolution probing higher redshifts. 

\vspace{-0.1in}
\begin{figure}[h!]
\centering
\includegraphics[width=0.48\linewidth]{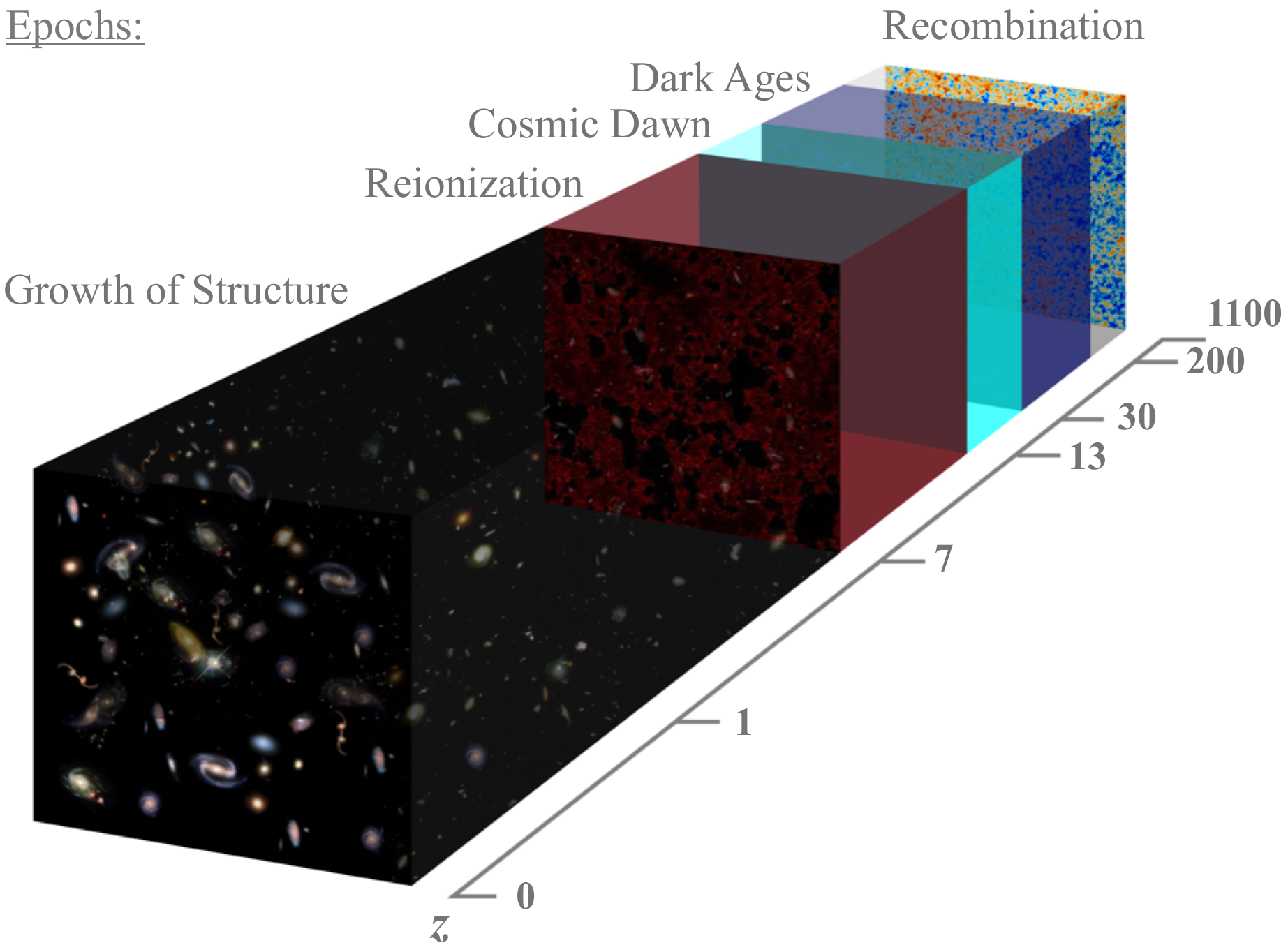}
\includegraphics[width=0.48\linewidth]{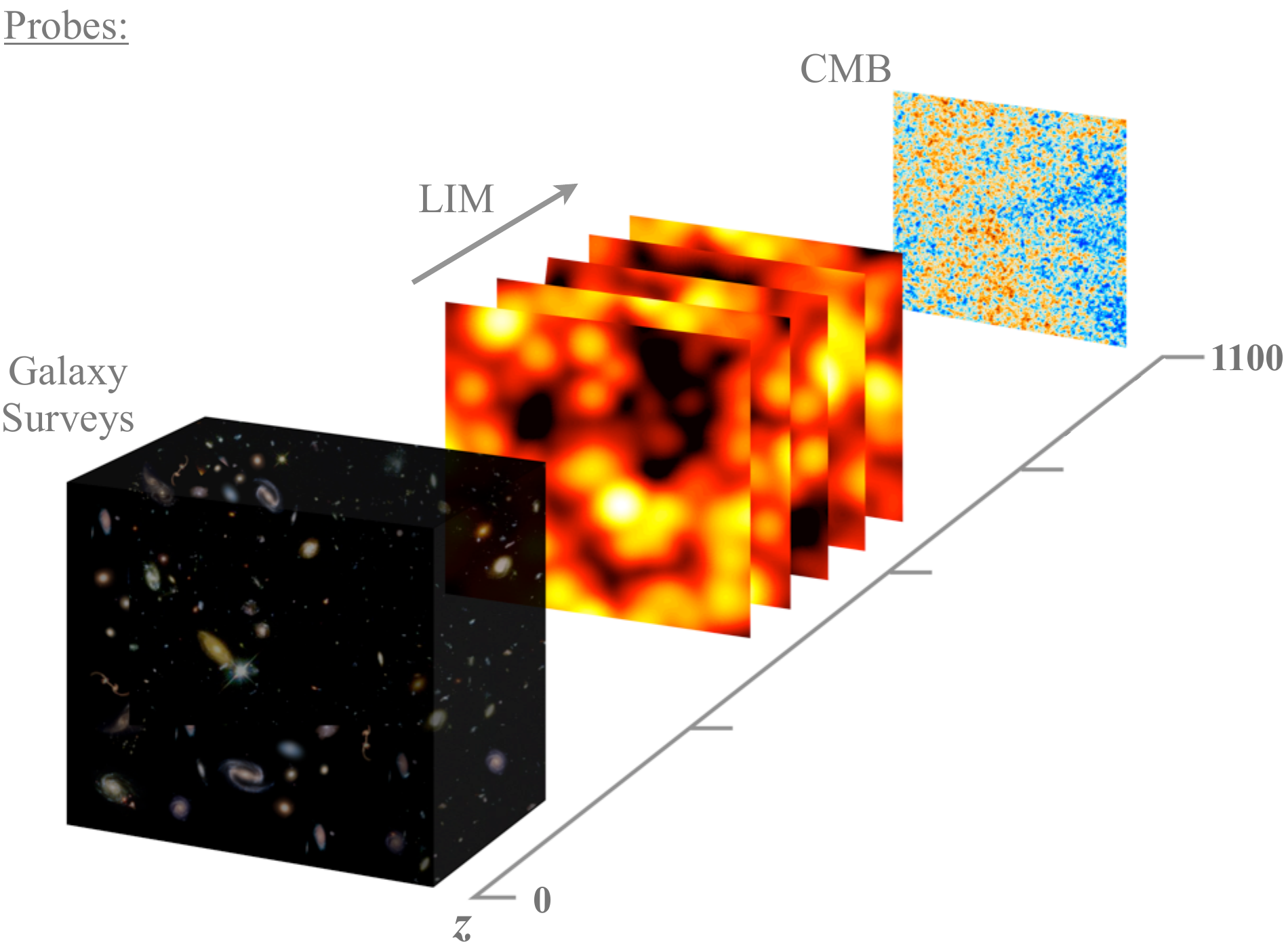}
\vspace{-0.1in}
\caption{Line-Intensity Mapping can access the uncharted $\gtrsim\!80\%$ volume of the observable Universe.}
\label{fig:IMVisual}
\end{figure}

Targets for LIM range from the 21-cm emission from neutral hydrogen in the IGM to line emission from galaxies, including the 21-cm line, as well as rotational carbon-monoxide (CO) transitions, the [CII] fine-structure line, the hydrogen Ly-$\alpha$ line, H-$\alpha$, H-$\beta$, [OII], [OIII], etc. The vast range of targeted wavelengths necessitates the employment of different instruments.

This paper describes the various science goals achievable by pushing LIM to its next frontier.
As we will stress throughout, there is unique potential in using multi-line observations, which motivates a coordinated effort to plan the future generation of LIM experiments.

\section{Galaxy Evolution}
\vspace{-0.2in}

Open questions abound surrounding galaxy evolution which can be uniquely studied with LIM.  How many stars were forming at any given time?  How do the various phases of the ISM evolve across cosmic time? How do galaxy properties vary with their large-scale environment, and how do they interact with the surrounding IGM? How do processes like supernova and AGN feedback shape the galaxies we see today? Through the combination of large spatial volumes, sensitivity to faint objects, and excellent redshift measurements, an ambitious LIM observational effort will provide a powerful complement to direct observations with future NASA flagship observatories such as JWST, WFIRST, HabEx, LUVOIR and Origins.

\underline{The Cosmic Star Formation history.} The star formation rate is a crucial probe of galaxy properties. However, measurements of the cosmic star formation rate density using single-object observations are highly uncertain at high redshifts, as only the very brightest galaxies can be detected by the required large-volume, multiwavelength galaxy surveys.  \cite{Kistler:2013jza}.  Conversely, surveys at low redshift can suffer from cosmic variance if they span small areas. Single object surveys may suffer additional selection bias if they do not blindly detect sources. {\em LIM yields a complete census of emitting gas that traces star formation across cosmic time.} Many of the lines to be targeted experimentally serve as excellent star formation tracers, including (but not limited to) CO \cite{Breysse2016A}, CII \cite{Yue2015}, and H$\alpha$ \cite{Silva2017}. 
By combining multiple tracers of cold gas and star formation, we can infer the star-formation rate, efficiency and possible evolution, as well as ISM properties, such as metallicity, temperature, and ionization state.

\underline{The High-Redshift ISM.}
A key advantage of LIM is that it can access multiple emission lines coming from different phases of the ISM and IGM.  CO lines trace the cold molecular clouds where stars are forming \cite{Bolatto2013}, while CII emission from those clouds and the surrounding photo-dissociation regions provide one of the most significant sources of cooling \cite{Dalgarno1972}. 21 cm, H-$\alpha$, and Ly-$\alpha$ probe hydrogen gas both in and around galaxies \cite{Villaescusa-Navarro2018,Silva2017,Pullen:2013dir}.  
[OIII], used by ALMA to great effect to measure redshifts of $z=7-9$ galaxies~\cite{Moriwaki2018}, can be used to study HII regions.  In the most distant sources, Pop III star formation can be observed in HeII \cite{2015MNRAS.450.2506V}, and molecular gas in the earliest, ultra-low-metallicity galaxies can be observed through the rotational transitions of HD \cite{Bromm2013}. Wide-field LIM measurements in the $100\!-\!600\,{\rm \mu m}$ range with Origins will open a new window for redshifted far-IR fine-structure transitions that dominate the ISM cooling in galaxies~\cite{Uzgil:2014pga,Serra:2016jzs}.
Meanwhile, molecular-line scan surveys with ALMA and the ngVLA can map the high-z volume density of cold star-forming gas~\cite{DominikWP}.

Ultimately, the real power of a multi-line experimental effort will come from  cross-correlation opportunities.  Correlations between different lines allow measurements of how e.g.\ CO and CII luminosities vary with respect to each other.  LIM also allows access to other lines which fall into target frequency bands and might be too faint to detect directly.  For example,  access to the $^{13}$CO isotopologue can provide an unprecedented understanding of the opacity and chemical enrichment of high-z molecular gas \cite{Breysse2016b,Zhang2018}. \emph{These opportunities motivate an effort to coordinate and share data between groups to maximize cross-correlation science output.}

\vspace{-0.25in}
\section{Reionization}
\vspace{-0.2in}

The epoch of reionization is an important, yet mostly unexplored period in the history of the universe when the first stars, accreting black holes, and galaxies formed, emitted ultraviolet light, and gradually photoionized neutral hydrogen gas in their surroundings \cite{Loeb13}. In current theoretical models, the IGM during reionization resembles a two-phase medium: ``bubbles'' of ionized gas form around the first luminous sources, while significantly neutral hydrogen regions remain intermixed. Some key, open questions which can be answered by LIM include: When did reionization occur? More precisely, what fraction of the IGM volume is in the ionized phase as a function of redshift? How did reionization proceed spatially; e.g., how large were the ionized bubbles at different stages of the reionization process? Answers to these questions will provide insights into the timing of early structure formation and the nature of the ionizing sources. Did star-forming galaxies or accreting black holes produce most of the ionizing photons? How did the first galaxies differ from subsequent generations? What are the thermal and chemical enrichment histories of the IGM? \emph{LIM can provide unique insight into the EoR by simultaneously mapping both the star forming galaxies which produce ionizing photons and the distribution of remaining neutral gas in the IGM.}

\underline{The Process of Reionization.}
LIM directly probes the process of reionization and the sources that drive it.
As is widely appreciated, the most direct probe of the IGM during reionization is the redshifted 21 cm line; the first detections of 21-cm fluctuations from the EoR are anticipated in the next decade \cite{DeBoer:2016tnn}.  Intensity maps of [CII], [OI], and [OIII] fine structure lines, CO rotational transitions, and H-$\alpha$ emission,  among others, may be used to trace the galaxy distribution in the same cosmological volume as the 21 cm observations. Mapping both the galaxies themselves and the surrounding neutral gas in the IGM will dramatically improve our understanding of the fundamental interplay between the ionizing sources and the IGM \cite{Lidz:2008ry,Gong:2011mf,Lidz:2011dx}.  In conjunction, Ly-$\alpha$ maps, which can be made using the recently-announced SPHEREx mission \citep{Dore:2014cca} probe some combination of the sources and the intergalactic gas; recombinations partly source this line emission in the ISM of high redshift galaxies and from recombinations in more diffuse intergalactic gas \cite{Pullen:2013dir,Silva13,Visbal:2018dsi}. The line photons subsequently scatter off of neutral hydrogen in the interstellar, circumgalactic, and intergalactic media. The Ly-$\alpha$ emission fluctuations are then modulated, in part, by the presence of ionized bubbles. Multi-line intensity mapping will probe a wide range of spatial scales and characterize the physical processes at play during reionization.

\vspace{-0.25in}
\section{Fundamental Cosmology}
\vspace{-0.2in}

Cosmology has recently entered a golden era. 
Precision observational cosmology, led by 
measurements of the CMB and augmented by input from galaxy surveys, has pinned down the parameters of the standard cosmological model ($\Lambda$CDM) to around 1\% uncertainty~\cite{Aghanim:2018eyx}. Several gaping holes remain in the theory, such as: How did inflation begin and come to an end? What makes up the dark matter in the Universe? What is the nature of dark energy? 
Going forward, to gain fresh insight into these fundamental questions, we must develop new ways to harvest larger volumes of the observable Universe. 

{\em LIM provides an ideal means to test ideas within and beyond} $\Lambda$CDM.  The emitting species observed by LIM surveys are biased tracers of the underlying dark matter density field, making them an excellent probe of large-scale structure.
Ambitious LIM experiments reveal invaluable new information by mapping significant parts of the sky over extended redshift epochs, and across a wide range of scales. In comparison, the potential of the CMB to constrain some of the simplest extensions of $\Lambda$CDM, such as primordial non-gaussianities and evolving dark energy, has been largely exhausted. Most existing galaxy redshift surveys have been restricted to $z<1$, and thus have covered only a small corner of the space of accessible linear modes. While future wide-field galaxy surveys will be able to probe much larger scales, they are much costlier, and naturally limited in sensitivity to faint sources.

\underline{Dark energy and modified gravity.}  LIM measurements of baryon acoustic oscillations can be made continuously from low to high redshifts, enabling a test of the time-variation of the dark-energy equation of state \citep{Dinda2018}. In particular, {\em LIM will shed light on the growing tension between local and CMB measurements of the Hubble parameter}~\cite{Bernal:2016gxb} by bridging the enormous distance gap between their sources. {\em LIM can also constrain modified gravity theories beyond their equation-of-state}~\cite{Carucci:2017cnn}, by testing various screening mechanisms, whose signatures are environment-dependent and differ between alternative theoretical manifestations~\cite{Jain:2013wgs,Hall2013,Pourtsidou2016}.  

\underline{Inflation.} Limits on primordial non-gaussianity from constraints on the scale-dependent bias in the two-point function, enabled by LIM measurements of line emission over extended redshift volumes at large scales, can potentially approach the coveted target of $\sigma(f_{\rm NL})\!\sim\!1$~\cite{MoradinezhadDizgah:2018zrs,Fonseca2018}.  {\em LIM may thus help distinguish between single and multi-field models of inflation}~\cite{dePutter:2016trg}. Unlike galaxy surveys, LIM also has the potential to go beyond this target (e.g.\ 21 cm~\cite{Munoz:2015eqa}).

\underline{Dark matter.} 
{\em LIM can potentially improve by as much as ten orders of magnitude the experimental sensitivity to radiative decays or annihilations from dark-matter particles}~\cite{Creque-Sarbinowski:2018ebl}, as the photons from monoenergetic decays will be correlated with the mass distribution, which can be determined from spectral intensity maps (and potentially their cross-correlation with galaxy or weak-lensing surveys). Access to the cosmic dawn era (when the first stars were born) via the 21-cm line can provide effective constraints on warm dark matter~\cite{Sitwell:2013fpa,Carucci:2015bra,Chatterjee:2019jts} and on models of dark-matter--baryon interactions~\cite{Tashiro:2014tsa,Munoz:2015bca,Barkana:2018lgd,Fialkov:2018xre,Munoz:2018jwq}, as this is the moment in the history of the Universe when the baryon temperature was closest to that of the {\it cold} dark matter.

\underline{Improving $\Lambda$CDM constraints.} 
Line emission is inherently tied to astrophysical processes, allowing LIM surveys to jointly constrain cosmology and astrophysics. Measurements of 21\,cm during reionization can improve our measurement of the optical depth to reionization~\cite{Liu:2015txa}, which limits fundamental cosmology from the CMB. Measurements of large-scale structure in the late universe could access  the BAO scale as well as the abundance of emitting gas, jointly conducting a survey for galaxy evolution and the expansion of the universe. 
Gravitational lensing of the observed line intensity fluctuations carries further information about low-redshift structure~\cite{Foreman:2018gnv} that will enhance many of the constraints listed above.

\vspace{-0.25in}
\section{Synergy with other Observations}
\vspace{-0.2in}

\underline{Deep galaxy observations.}
Observatories like ALMA, the ngVLA and the soon-to-be-launched JWST will probe the detailed properties of handfuls of bright galaxies, while LIM surveys will make it possible to determine if these properties hold across the cumulative distribution of fainter galaxies. This sensitivity is of particular advantage during the EoR, as present measurements suggest that low luminosity galaxies produce most of the ionizing photons \cite{Robertson:2015uda}  and these sources may remain undetectable even in deep JWST observations.  

\underline{Large galaxy surveys.}
Upcoming wide surveys such as WFIRST and LSST will yield large galaxy catalogs ideal for cross-correlation with intensity maps. Similar cross-correlations have already aided in measuring diffuse line intensities to probe the ISM by removing uncorrelated foregrounds such as the Milky Way, and this will only continue with more sensitive instruments. In addition, cross power spectra allow the determination of line luminosities of hundreds or thousands of galaxies at once, a measurement typically unfeasible with directed observations \cite{Wolz:2015ckn}. {\em These cross-correlations can provide critical insights into astrophysical mysteries ranging from conditions in diffuse IGM \cite{Croft:2018rwv} to the makeup of dark matter \cite{Creque-Sarbinowski:2018ebl}}.  LIM can also improve the science yield of imaging surveys by independently measuring the redshift distributions of their targets, e.g.\ via clustering-based redshift estimation \cite{Menard:2013aaa,Cunnington2019}. Lastly, combining LSS and LIM affords a multi-tracer approach, mitigating  cosmic variance.

\underline{Ly-$\alpha$ forest.}
  DESI~\cite{Aghamousa:2016zmz} and WEAVE~\cite{Pieri:2016jwo} will allow  Ly-$\alpha$ forest measurements with improved spatial resolution, area and data quality. Cross-correlation with Ly-$\alpha$ LIM will help understand the impact of Ly-$\alpha$ emitters on their environment and boost BAO measurements~\cite{Croft:2018rwv}.

\underline{CMB.} 
LIM-CMB correlations~\cite{Ballardini:2018cho} can retrieve redshift information for secondary CMB anisotropies, such as lensing, as well as hot gas tracers like the thermal and kinetic Sunyaev-Zel'dovich perturbations, which LIM can complement by mapping the cold gas distribution.

\underline{Foreground Rejection.} 
Foreground contamination impacts inferences from intensity mapping of a single line.  Though methods to remove these foregrounds exist~\cite{Kovetz:2017agg}, cross-correlations with other large-scale structure tracers will help ensure the robustness of LIM observations.

\underline{Discovery space.}
Since LIM uses new observational approaches and technology that cast a wide net across the universe,  it is also likely to discover or characterize new phenomena. A recent example is the CHIME detection of fast radio bursts at frequencies $\gtrsim400\,{\rm MHz}$~\cite{Amiri:2019qbv}.

\vspace{-0.25in}
\section{Outlook and Summary}
\vspace{-0.2in}

Currently, there are initial LIM detections at relatively low redshifts in 21 cm ($z\!\sim\! 0.8$) \cite{Masui13,Anderson2018}, [CII] ($z\! \sim\! 2.5$) \cite{Pullen:2017ogs}, and Ly-$\alpha$ ($z\! \sim\! 3$) \cite{Croft:2015nna,Croft:2018rwv} through cross-correlations with traditional galaxy or quasar surveys, as well as a CO auto-spectrum detection ($z \sim 3$) in the shot-noise regime \cite{Keating2016}. Early in the coming decade, many efforts are pushing to strengthen the statistical significance of these detections and to increase the coverage in redshift and  volume. 
Further efforts are in
the planning phases, promising to advance early measurements beyond the level of mere detections, to achieve detailed characterizations of the LIM signals.

\vspace{-0.1in}

\begin{figure}[h!]
\centering
\includegraphics[width=\linewidth]{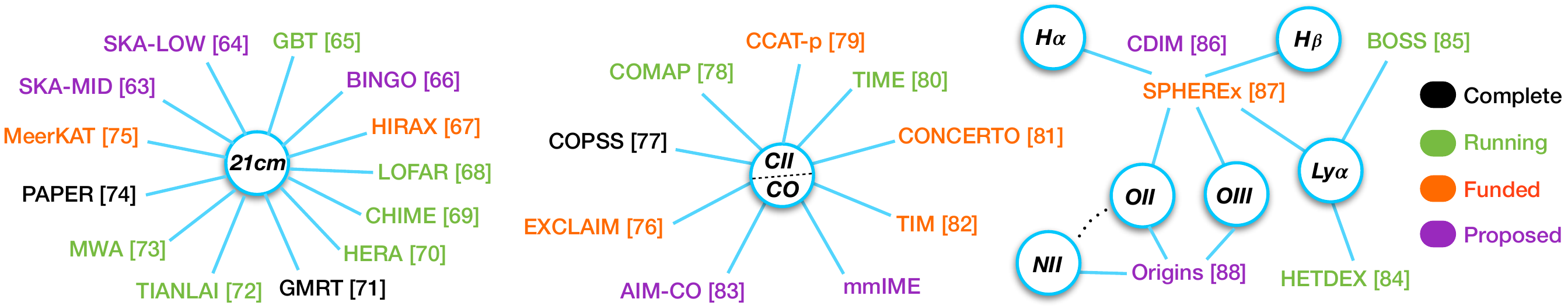}
\vspace{-0.3in}
\caption{\!\!\!  Various current, upcoming and future line-intensity mapping instruments~\cite{Bacon:2018dui,Koopmans:2015sua,Masui:2012zc,2013MNRAS.434.1239B,Newburgh:2016mwi,2013A&A...556A...2V,2014SPIE.9145E..22B,deboer_2017,Pen:2008ut,Wu:2016vzu,2013PASA...30....7T,Parsons:2013dwa, Santos:2017qgq,Padmanabhan:2018yul,2015ApJ...814..140K,Li:2015gqa,Stacey:2018yqe,Crites:2014,2018A&A...609A.130L,2018AAS...23132804A,2016AAS...22742604B,Hill:2008mv,Gil-Marin:2014sta,Cooray:2016hro,2014arXiv1412.4872D,2018arXiv180909702T}.}
\label{fig:IMVisual}
\end{figure}

\vspace{-0.15in}

In conjunction with the upcoming surveys, it will be important to refine theoretical modeling efforts. New multi-scale simulation models are required to best capture the enormous range in spatial scale relevant for line-intensity mapping observations, which involve the interstellar media of individual dwarf galaxies out to $\sim\!{\rm Gpc}$ cosmological length scales \cite{Villaescusa-Navarro2018,Stein2019}. In addition, targeted observations of individual galaxies over cosmic time will help in calibrating scaling relations and aid the interpretation of upcoming LIM measurements.

LIM is uniquely poised to address a broad range of science goals, from the history of star formation and galaxy evolution, through the details of the epoch of reionization, to deeper insight into the critical questions of fundamental cosmology. 
{\em This motivates an active research program over the coming decade, including continued investments in multiple line-intensity mapping experiments  to span overlapping cosmological volumes, along with  support for simulation and modeling efforts.}

\newpage

{\bf This white paper has been endorsed by:}

Muntazir Abidi$^{1}$, 
James Aguirre$^{2}$, 
Yacine Ali-Ha\"imoud$^{54}$,
David Alonso$^{3}$, 
Marcelo Alvarez$^{60}$,
Jacobo Asorey$^{4}$, 
Mario Ballardini$^{5}$, 
Kevin Bandura$^{6,7}$, 
Nicholas Battaglia$^{8}$, 
Daniel Baumann$^{9,10}$, 
Chetan bavdhankar$^{11}$, 
Angus Beane$^{2}$,
Adam Beardsley$^{12}$, 
Charles Bennett$^{13}$, 
Bradford Benson$^{14,15}$, 
Jos\'{e} Luis Bernal$^{16,17}$, 
Florian Beutler$^{18}$, 
J.~Richard Bond$^{20}$,
Julian Borrill$^{19}$, 
Geoffrey Bower$^{63}$,
Patrick C.~Breysse$^{20}$, 
Philip	Bull$^{72}$,
Zheng Cai$^{21}$, 
Isabella P.~Carucci$^{67}$,
Francisco J.~Castander$^{22}$, 
Emanuele Castorina$^{23}$, 
Jon\'{a}s Chaves-Montero$^{24}$, 
Xuelei Chen$^{52}$,
Dongwoo Chung$^{28}$,
Sarah Church$^{61}$,
Kieran Cleary$^{48}$,
J.~D.~Cohn$^{25}$, 
Asantha Cooray$^{57}$,
Thomas Crawford$^{26,15}$, 
Abby	Crites$^{48}$,
Rupert A.~C.~Croft$^{27}$, 
Guido D'Amico$^{28}$, 
Josh Dillon$^{60}$,
Olivier Dor\'e$^{68}$,
Kelly A.~Douglass$^{29}$, 
Cora Dvorkin$^{30}$, 
Thomas Essinger-Hileman$^{31}$, 
Andrea Ferrara$^{64}$,
Simone Ferraro$^{19}$,
Jos\'e Fonseca$^{70}$,
Simon Foreman$^{20}$, 
Steve Furlanetto$^{62}$,
Martina Gerbino$^{24}$, 
Vera Gluscevic$^{32}$, 
Alma X. Gonz\'alez-Morales$^{33}$, 
Christopher M.~Hirata$^{34}$, 
Ilian T.~Iliev$^{35}$, 
Danny Jacobs$^{12}$, 
Matthew C.~Johnson$^{36,37}$, 
Marc Kamionkowski$^{13}$, 
Kirit	Karkare$^{15}$,
Jean-Paul Kneib$^{38}$, 
Ely D.~Kovetz$^{39}$, 
Chien-Hsiu Lee$^{40}$, 
Guilaine Lagache$^{56}$,
Adam Lidz$^{2}$,
Adrian Liu$^{41}$,
Yi Mao$^{42}$, 
Dan Marrone$^{69}$,
Kiyoshi Masui$^{43}$, 
P.~Daniel Meerburg$^{44,1,45}$, 
Raul Monsalve$^{41}$,
Julian Mu\~noz$^{30}$,
Michael D.~Niemack$^{8}$, 
Paul O'Connor$^{46}$, 
Hamsa Padmanabhan$^{20}$,
Matthew M.~Pieri$^{47}$, 
Abhishek Prakash$^{48}$, 
Anthony Pullen$^{54}$,
Mubdi Rahman$^{58}$,
Graziano Rossi$^{49}$, 
Shun Saito$^{55}$,
Emmanuel Schaan$^{19}$,
Neelima Sehgal$^{50}$, 
Hee-Jong Seo$^{51}$, 
Marta Silva$^{65}$,
Feng Shi$^{4}$, 
An\v{z}e Slosar$^{46}$,
Rachel	Somerville$^{71}$,
David Spergel$^{59}$,
Srivatsan Sridhar$^{4}$, 
Gordon Stacey$^{8}$,
Albert Stebbins$^{14}$,
George Stein$^{20}$,
Eric R. Switzer$^{31}$, 
Peter Timbie$^{64}$,
Cora Uhlemann$^{1}$, 
Eli	Visbal$^{71}$,
Fabian Walter$^{53}$,
Martin White$^{60}$,
Gong-Bo Zhao$^{52,18}$

\newpage

\bibliography{ScienceReport}
\bibliographystyle{arxiv}

\newpage


 $^{1}$ \damtp \\
$^{2}$ \UPenn \\
$^{3}$ \Oxford \\
$^{4}$ \KASSI \\
$^{5}$ \UWC \\
$^{6}$ \WVU \\
$^{7}$ \WVUGWAC \\
$^{8}$ \Cornell \\
$^{9}$ \GRAPPA \\
$^{10}$ \ITFA \\
$^{11}$ \NCBJ \\
$^{12}$ \ASU \\
$^{13}$ \JHU \\
$^{14}$ \FNAL \\
$^{15}$ \KICP \\
$^{16}$ \ICC \\
$^{17}$ \FQAUB \\
$^{18}$ \Port \\
$^{19}$ \LBL \\
$^{20}$ \CITA \\
$^{21}$ \SCIPP \\
$^{22}$ \ICE \\
$^{23}$ \UCBP \\
$^{24}$ \ANLHEP \\
$^{25}$ \UCBSSL \\
$^{26}$ \UChicago \\
$^{27}$ \CMUCosmo \\
$^{28}$ \Stanford \\
$^{29}$ \UR \\
$^{30}$ \HarvardPhys \\
$^{31}$ \GSFC \\
$^{32}$ \UFL \\
$^{33}$ \UGTO \\
$^{34}$ \OSU \\
$^{35}$ \SussexAstronomy \\
$^{36}$ \PI \\
$^{37}$ \YorkU \\
$^{38}$ \EPFL \\
$^{39}$ \BenGurion \\
$^{40}$ \NOAO \\
$^{41}$ \McGill \\
$^{42}$ \Tsinghua \\
$^{43}$ \MIT \\
$^{44}$ \kavli \\
$^{45}$ \VSI \\
$^{46}$ \BNL \\
$^{47}$ \CPPM \\
$^{48}$ \Caltech \\
$^{49}$ \Sejong \\
$^{50}$ \StonyBrook \\
$^{51}$ \OU \\
$^{52}$ \NAOC \\
$^{53}$ \Heidelberg \\
$^{54}$ \NYU \\
$^{55}$ \Missouri \\
$^{56}$ \Marseille \\
$^{57}$ \UCI \\
$^{58}$ \dunlap \\
$^{59}$ \Princeton \\
$^{60}$ \UCB \\
$^{61}$ \KIPAC \\
$^{62}$ \UCLA \\
$^{63}$ \Sinica \\
$^{64}$ \UWMadison \\
$^{65}$ \Pisa \\
$^{66}$ \Oslo \\
$^{67}$ \UCL \\
$^{68}$ \JPL \\
$^{69}$ \UAS \\
$^{70}$ \UNIPD \\
$^{71}$ \CCA \\
$^{72}$ \QMUL 

\end{document}